\documentclass{xray_symp}
\usepackage{graphicx}
\usepackage{natbib}

\newcommand{\eqref}[1]{eq.~(\ref{#1})}
\newcommand{\keV}{\mbox{keV}}

\newcommand{\Ginga}{\textsl{Ginga}\ }
\newcommand{\lmcone}{LMC~X-1\ }
\newcommand{\lmcthree}{LMC~X-3\ }

\newcommand{\Msun}{\mbox{M}_\odot }

\begin{document}

\title{RXTE Observations of LMC~X-1 and LMC~X-3}
\author{J.~Wilms\inst{1} \and M.A.~Nowak\inst{2} \and J.B.~Dove\inst{3},
  K.~Pottschmidt\inst{1} \and
  W.A.~Heindl\inst{4} \and M.C.~Begelman\inst{2} \and R.~Staubert\inst{1}} 
\authorrunning{J. Wilms et al.} 
\institute{%
Institut f\"ur Astronomie und Astrophysik -- Astronomie,
Waldh\"auser Str. 64, D-72076 T\"ubingen, Germany
\and
JILA, University of Colorado, Boulder, CO 80309-440, U.S.A.
\and
CASA, University of Colorado, Boulder, CO 80309-389, U.S.A.
\and
CASS, University of California, San Diego, La Jolla, CA 92093, U.S.A.}

\thesaurus{XXXXXXXX}  
\date{X}
\maketitle

\begin{abstract}  
  Of all known persistent stellar-mass black hole candidates, only \lmcone
  and \lmcthree consistently show spectra that are dominated by a soft,
  thermal component.  We present results from long (170\,ksec) Rossi
  \mbox{X-ray} Timing Explorer (RXTE) observations of \lmcone and \lmcthree
  made in 1996~December.  The spectra can be described by a multicolor disk
  blackbody plus an additional high-energy power-law.  Even though the
  spectra are very soft ($\Gamma\sim 2.5$), RXTE detected a significant
  signal from \lmcthree up to energies of 50\,keV, the hardest energy at
  which the object was ever detected.
  
  Focusing on \lmcthree, we present results from the first year of an
  ongoing monitoring campaign with RXTE which started in 1997 January.
  We show that the appearance of the object changes considerably over
  its $\sim$200\,d long cycle.  This variability can either be
  explained by periodic changes in the mass transfer rate or by a
  precessing accretion disk analogous to Her~X-1.
\end{abstract}

\section{Introduction}
Since the discovery of Cygnus~X-1 in 1964 \citep{bowyer:65a}, the
study of galactic black holes (BHs) has shown that these objects
reveal a large variety of states, which are characterized by their
distinct spectral shapes and temporal behaviors. The most important
states which have been identified are the so called ``hard state'',
which is characterized by a hard X-ray spectrum with a photon index
$\Gamma=1.7$ and large variability (${\rm rms}=30\%$, see
contributions by Pottschmidt et al.\ and Belloni in this volume), and
the ``soft state'', which is spectrally softer ($\Gamma \sim 2.5$) and
characterized by less variability. The overall luminosity of sources
in the soft state appears to be higher than that of sources in the
hard state \citep{nowak:95a}.

Despite the fact that the soft state is very common in galactic BHs,
most observational attention has been concentrated on the hard state,
since most of the brighter galactic BHs are found in this state and
only show occasional state switches to the soft state.  Only two of
the persistent galactic BHs, LMC X-1 and LMC X-3, are always found in
the soft state.  These objects were discovered by \textsl{Uhuru}
during scans of the Large Magellanic Cloud \citep{leong:71a}.
Although their intrinsic luminosity is quite high (a few $10^{38}\,\rm
erg/sec$), the large distance of the LMC prevented the detailed study of
these objects for a long time. Such a study became feasible with the
advent of detectors with large effective areas.  \Ginga results on
\lmcone and \lmcthree revealed that both sources exhibit very
interesting physical behavior, such as a possible low-frequency QPO
and long term spectral variability \citep{ebisawa:91a,ebisawa:93a}.
No systematic monitoring was done by \textsl{Ginga}, however, and the
absence of an instrument sensitive above $\sim$20\,\keV\ prohibited
gathering information about the high energy spectrum.

To enable a systematic study of the soft state we have monitored \lmcone
and \lmcthree with the Rossi X-ray Timing Explorer (RXTE) since the end of
1996 in three to four weekly intervals, which has provided an unique
opportunity to study their long term behavior. To facilitate the
understanding of the spectrum, the campaign started with 170\,ksec long
observations of both sources. In this contribution we present first results
from the spectral analysis of the long observations.  Note that the results
presented here are preliminary since the campaign will be continued
throughout all of 1998.  We start, in Sect.~\ref{sec:rxte}, with a
description of our data analysis methodology. We then present the outcome
of the analysis of the long observations of \lmcone
(Sect.~\ref{sec:lmcone}) and \lmcthree (Sect.~\ref{sec:lmcx3long}). The
spectral variability of \lmcthree during the first year of the campaign is
described in Sect.~\ref{sec:threemonit}.

\section{Data Analysis}\label{sec:rxte}
The data presented here were obtained with the Rossi X-ray Timing Explorer
(RXTE). Onboard RXTE are two pointed instruments, the Proportional Counter
Array (PCA) and the High Energy X-ray Timing Experiment (HEXTE), as well as
the All Sky Monitor (ASM). We used the standard RXTE data analysis
software, ftools~3.5. Spectral modeling was done using XSPEC, version
10.00s \citep{arnaud:96a}. In the meantime a revised release of the data
analysis software and of the response matrices has become available. The
results of an analysis using these improved tools are the subject of a
forthcoming paper \citep{wilms:98b}.

The PCA consists of five co-aligned Xenon proportional counter units
(PCUs) with a total effective area of about $6500\,\mbox{cm}^2$. The
instrument is sensitive in the energy range from 2\,keV to $\sim
60$\,keV \citep{jahoda:96b}, although response matrix uncertainties
currently limit the usable energy range to 2.5--30\,keV. We used a
pre-release version of the PCA response matrices, version~2.2.1, for
the spectral analysis (Jahoda, 1997, priv.\ comm.). The spectral
calibration of the instrument appears to be understood on the 2\%
level for this version of the matrix.  Even for count-rates as small
as those of \lmcone and \mbox{LMC X-3}, the systematic uncertainty
affects the data. Therefore, we added a 2\% systematic error to the
data.  See \citet{dove:97c} and \citet{wilms:98c} for an in-depth
discussion of the PCA calibration issues. Background subtraction of
the PCA data was performed analogously to our previous study of Cyg
X-1 \citep{dove:97c}. Since the major uncertainty of the PCA
background model is in the description of the radioactivity induced by
spallation of the detector material during passages of the satellite
through the South Atlantic Anomaly (SAA), we ignored data measured in
the 30\,minutes after SAA passages.

HEXTE consists of two clusters of four NaI/CsI-phoswich scintillation
counters that are sensitive from 15 to 250\,keV. A full description of
the instrument is given by \citet{rothschild:98a}. Background
subtraction is done by source-background switching. We used the
standard HEXTE response matrices of 1997~March 20. Data measured above
20\,keV were used.  In our spectral fits we took care of the
intercalibration of the instruments by introducing a multiplicative
constant.  We consistently found the HEXTE fluxes to be 75\% of the
PCA fluxes. Most probably, this offset between fluxes is due to a
slight misalignment of the HEXTE honeycomb collimators (Heindl, 1998,
priv.\ comm.).

\section{LMC X-1: The Long Observation}\label{sec:lmcone}
LMC X-1 is a good candidate for a black hole.  Using a large number of
\textsl{ROSAT} HRI observations, \citet{cowley:95a} were able to
identify the counterpart with ``star number 32'' of
\citet{cowley:78a}.  This object has a mass function of only
$f=0.144\,\Msun$, but including other evidence the mass of the compact
object appears to be $M>4\,\Msun$ \citep{hutchings:87a}. The
luminosity of the object is about $2\times 10^{38}\,\rm erg/s$
\citep{long:81a} and was not found to be variable \citep{sunyaev:90a}.
We could verify the latter statement in our monitoring campaign. We
therefore concentrate here on the results from the long RXTE
observation from 1996~December 6 to 8, which is a typical example for
the spectrum of LMC X-1. The object was circumpolar during the three
days of the observation. We only use data from time intervals when all
five PCA-PCUs were turned on.  Taking also our 30\,minutes SAA
exclusion time interval into account, a total of 80\,ksec of data were
obtained.

\begin{figure}
\resizebox{\hsize}{!}{\includegraphics{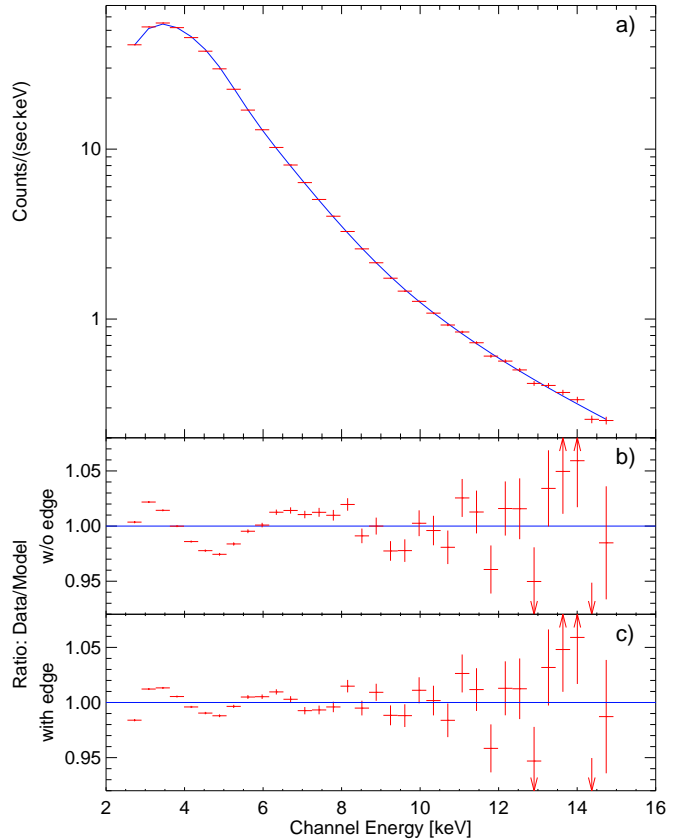}}
\caption{\textbf{a} PCA spectrum of 80\,ksec of on-source data on LMC X-1.
  \textbf{b} Ratio between the data and the best fit spectral model without
  and \textbf{c} with a smeared edge feature.}\label{fig:lmcx1spec}
\end{figure}

In Fig.~\ref{fig:lmcx1spec} we show the total PCA spectrum measured
during that time.  The spectrum can be well described by either a pure
black-body or a multicolor disk blackbody (MCD) with
$kT=1^{+0.02}_{-0.04}$\,keV to which a high energy power-law with a
photon index $\Gamma=3.65^{+0.05}_{-0.07}$ is added.  In both cases,
$\chi^2/{\rm dof}=50.5/30$.  The multicolor disk black body is an
approximation to the spectrum of an accretion disk with $T(r)\propto
r^{-3/4}$, i.e., a simple $\alpha$-disk.  See
\citet[eq.~4]{mitsuda:84a} for a detailed description of this model.
The parameters are consistent with those found in previous
investigations \citep[and references
therein]{schlegel:94a,ebisawa:91a}.  Introduction of a smeared edge in
the region around 7.5\,keV, as was required in the \textsl{Ginga} and
\textsl{BBXRT} analysis, slightly increases the quality of the fit
(Fig.\ \ref{fig:lmcx1spec}b and c), although the uncertainty of the
PCA response matrix prohibits any statement whether the improvement is
real. Note, however, that in this energy region the soft disk
radiation and the power-law have the same strength, so that the edge
might be just a feature associated with this region of overlap.

\begin{figure}
\resizebox{\hsize}{!}{\includegraphics{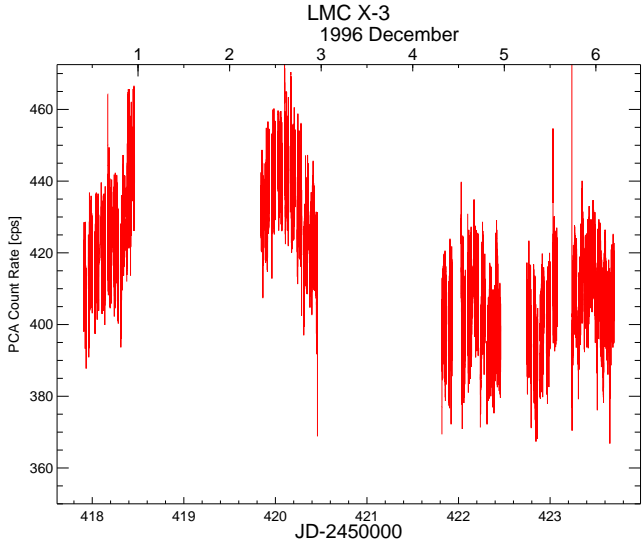}}
\caption{Background-subtracted light-curve of LMC~X-3 for the long
  observation in 1996 December. The short gaps are due to SAA passages and
  Earth occultations. The ``spikes'' are due to the PCA background model
  which is insufficient directly after the SAA passage (these data were
  excluded in the analysis). Note the ``flare'' in the first two data
  blocks.}\label{fig:lmcx3long}
\end{figure}

\section{LMC X-3}\label{sec:lmcthree}
\subsection{The Long Observation}\label{sec:lmcx3long}
\lmcthree is the most luminous BH in the LMC. The object has a peak
X-ray luminosity of about $4\times 30^{38}$\,erg/s and is variable by
a factor of about four on time scales of 100\,d or 200\,d \citep[see
also Sect.~\ref{sec:threemonit}]{cowley:91a}.  The optical
counterpart is a well established B3\,V star in a 1.7\,d orbit. The
mass function of the system is $f=2.3\,\Msun$ \citep{cowley:83a}.
Using the absence of X-ray eclipses to determine an upper limit for
the inclination the mass of the compact object is found to be above
9\,$\Msun$ \citep{cowley:83a} and therefore a very safe candidate for
a Black Hole.

\begin{figure}
\resizebox{\hsize}{!}{\includegraphics{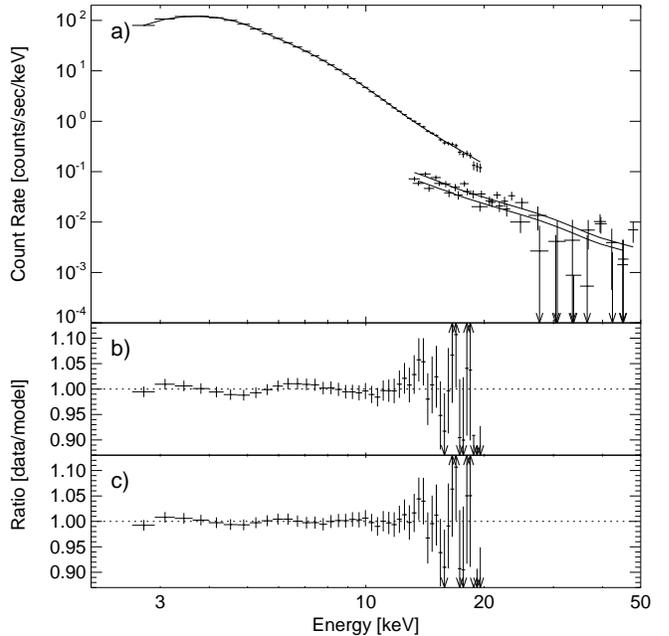}}
\caption{Combined PCA/HEXTE spectrum of the 
  long observation of LMC~X-3 (including the ``flare'' -- we did not find
  any variation in spectral shape over the whole observation).}
  \label{fig:lmcx3spec} 
\end{figure}

In Fig.~\ref{fig:lmcx3long} we show the lightcurve of our \mbox{LMC X-3}
observation. The larger countrate of the object compared to LMC X-1 allows
the inclusion of the high energy HEXTE data. Fig. \ref{fig:lmcx3spec} shows
that the object is detected out to 50 keV, the highest energy at which LMC
X-3 has ever been observed. The joint PCA/HEXTE data can be well described
by a multicolor disk black-body with $kT_{\rm in}=1.25\pm 0.01\,\keV$ plus
a power-law with a photon-index of $\Gamma=2.5\pm 0.2$ ($\chi^2/{\rm
  dof}=114/117$). These values are in agreement with previous observations
by \textsl{EXOSAT} and \Ginga \citep{treves:88a,treves:90a}.  Adding a
smeared edge at 7.5\,keV to the data improves the fit (Figs.\ 
\ref{fig:lmcx3spec}b and~c), but as with LMC X-1 the feature might be
caused by the transition between the disk black-body and the
power-law, and may not be a true spectral feature in its own right.

\subsection{Spectral variability}
\label{sec:threemonit}

\begin{figure}
\resizebox{\hsize}{!}{\includegraphics{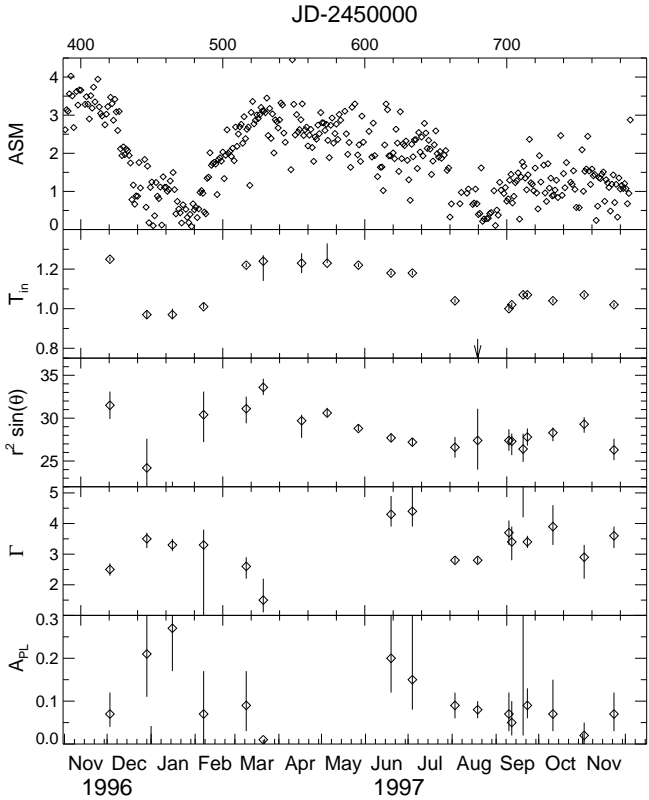}}
\caption{\label{fig:lmcx3asm}\textbf{Top:} RXTE/ASM soft
  X-ray flux, illustrating the long term variability. \textbf{Following
  Plots:} Temporal variability of the best-fit parameters for the first 20
  monitoring observations, $T_{\rm in}$: inner disk temperature in the
  multicolor disk model (keV), $r^2\sin\theta$: normalization of the MCD
  model, $\Gamma$: photon index of the high energy power-law, $A_{\rm PL}$:
  Normalization of the power-law (photon flux at 1\,keV). }
\end{figure}

Since the MCD plus power-law model was shown to give a good
description of the long observation we also used this model to
describe the data from our short (10\,ksec) monitoring observations.
In Fig.~\ref{fig:lmcx3asm} we present the results of this modeling for
the first year of our campaign.  Note that all caveats associated with
our use of the older background model also apply to these data
\citep[cf. ]{dove:97c}. In our fits we find that for the first half of
the observations, until about 1997~May, lower MCD temperatures are
correlated with softer high energy power-laws
(Fig.~\ref{fig:lmcx3asm}). Such a tendency appears not to be present
in the second half of the observations.  This could indicate that the
soft and hard spectral components, which we associate with the
accretion disk and a Comptonizing corona as most plausible places of
origin, are produced in geometrically separate regions of the system.
We will be able to test this claim with the data from the second year
of observations which are currently being measured.

\begin{figure}
\resizebox{\hsize}{!}{\includegraphics{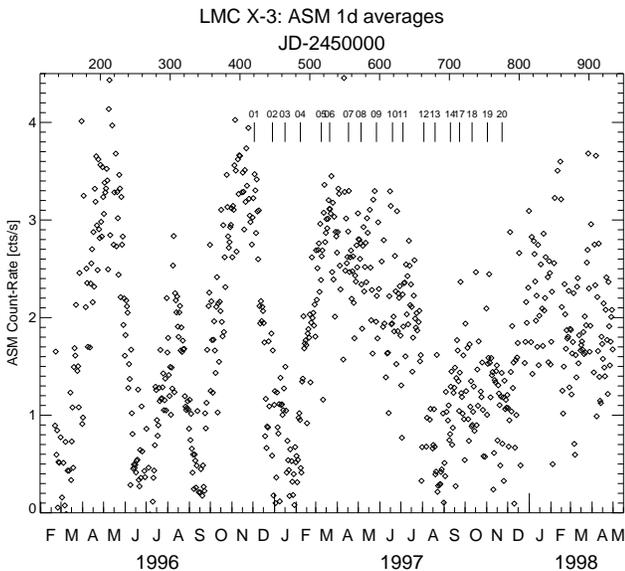}}
\caption{RXTE ASM lightcurve since the launch of RXTE.}\label{fig:lmcx3monit}
\end{figure}

In Fig. \ref{fig:lmcx3monit} we place the above results in the bigger
picture resulting from the RXTE ASM data. The ASM data for 1996
clearly show the pure $\sim 200$\,d periodicity found by
\citet{cowley:91a}. After that year, however, the sinusoidal variation
is replaced by a much more complicated variability pattern. Note that
at the same time the apparent correlation between the spectral
parameters shown in Fig.~\ref{fig:lmcx3asm} vanishes.  Since the
variability is either interpreted by a radiatively warped accretion
disk \citep{maloney:96a} or by a disk warped by an accretion disk wind
\citep[analogous to Her~X-1; ][]{schandl:96a}, the change in the
variability pattern could indicate that the accretion disk of LMC X-3
is unstable.  Further data, as well as the analysis of the data from
our current monitoring campaign, might clarify this phenomenon.

\acknowledgements The research presented in this paper has been financed by
NASA grants NAG5-3225, NAG5-4621, NAG5-4737, and a travel grant to J.W.
and K.P. from the DAAD.


\begin{thebibliography}{}

\bibitem[\protect\astroncite{Arnaud}{1996}]{arnaud:96a}
Arnaud K.A.,  1996,
\newblock In: Jacoby J.H., Barnes J. (eds.) Astronomical Data Analysis Software
  and Systems {V}. Astron.\ Soc.\ Pacific, Conf.\ Ser. 101, Astron.\ Soc.\
  Pacific, San Francisco, p.~17

\bibitem[\protect\astroncite{Bowyer et~al.}{1965}]{bowyer:65a}
Bowyer S., Byram E.T., Chubb T.A., Friedman H.,  1965, Science 147, 394

\bibitem[\protect\astroncite{Cowley et~al.}{1978}]{cowley:78a}
Cowley A.P., Crampton D., Hutchings J.B.,  1978, AJ 83, 1619

\bibitem[\protect\astroncite{Cowley et~al.}{1983}]{cowley:83a}
Cowley A.P., Crampton D., Hutchings J.B., et~al., 1983, ApJ 272, 118

\bibitem[\protect\astroncite{Cowley et~al.}{1995}]{cowley:95a}
Cowley A.P., Schmidtke P.C., Anderson A.L., {McGrath} T.K.,  1995, PASP 107,
  145

\bibitem[\protect\astroncite{Cowley et~al.}{1991}]{cowley:91a}
Cowley A.P., Schmidtke P.C., Ebisawa K., et~al., 1991, ApJ 381, 526

\bibitem[\protect\astroncite{Dove et~al.}{1998}]{dove:97c}
Dove J.B., Wilms J., Nowak M.A., et~al., 1998, MNRAS 289, 729

\bibitem[\protect\astroncite{Ebisawa et~al.}{1993}]{ebisawa:93a}
Ebisawa K., Makino F., Mitsuda K., et~al., 1993, ApJ 403, 684

\bibitem[\protect\astroncite{Ebisawa et~al.}{1991}]{ebisawa:91a}
Ebisawa K., Mitsuda K., Hanawa T.,  1991, ApJ 367, 213

\bibitem[\protect\astroncite{Hutchings et~al.}{1987}]{hutchings:87a}
Hutchings J.B., Crampton D., Cowley A.P., et~al., 1987, AJ 94, 340

\bibitem[\protect\astroncite{Jahoda et~al.}{1997}]{jahoda:96b}
Jahoda K., Swank J.H., Giles A.B., et~al., 1997,
\newblock In: Siegmund O.H. (ed.) {EUV}, X-Ray, and Gamma-Ray Instrumentation
  for Astronomy {VII}. Proc.\ SPIE 2808, SPIE, Bellingham, WA, p.59

\bibitem[\protect\astroncite{Leong et~al.}{1971}]{leong:71a}
Leong C., Kellogg K., Gursky H., et~al., 1971, ApJ 170, L67

\bibitem[\protect\astroncite{Long et~al.}{1981}]{long:81a}
Long K.S., Helfand D.J., Grabelsky D.A.,  1981, ApJ 248, 925

\bibitem[\protect\astroncite{Maloney et~al.}{1996}]{maloney:96a}
Maloney P.R., Begelman M.C., Pringle J.E.,  1996, ApJ 472, 582

\bibitem[\protect\astroncite{Mitsuda et~al.}{1984}]{mitsuda:84a}
Mitsuda K., Inoue H., Koyama K., et~al., 1984, PASJ 36, 741

\bibitem[\protect\astroncite{Nowak}{1995}]{nowak:95a}
Nowak M.A.,  1995, PASP 107, 1207

\bibitem[\protect\astroncite{Rothschild et~al.}{1998}]{rothschild:98a}
Rothschild R.E., Blanco P.R., Gruber D.E., et~al., 1998, ApJ 496, 538

\bibitem[\protect\astroncite{Schandl}{1996}]{schandl:96a}
Schandl S.,  1996, A\&A 307, 95

\bibitem[\protect\astroncite{Schlegel et~al.}{1994}]{schlegel:94a}
Schlegel E.M., Marshall F.E., Mushotzky R.F., et~al., 1994, ApJ 422, 243

\bibitem[\protect\astroncite{Syunyaev et~al.}{1990}]{sunyaev:90a}
Syunyaev R.A., {Gil'fanov} M., Churazov E., et~al., 1990, SvA Letters 16, 55

\bibitem[\protect\astroncite{Treves et~al.}{1988}]{treves:88a}
Treves A., Belloni T., Chiapetti L., et~al., 1988, ApJ 325, 119

\bibitem[\protect\astroncite{Treves et~al.}{1990}]{treves:90a}
Treves A., Belloni T., Corbet R.H.D., et~al., 1990, ApJ 364, 266

\bibitem[\protect\astroncite{Wilms et~al.}{1998a}]{wilms:98c}
Wilms J., Nowak M.A., Dove J.B., et~al., 1998a, ApJ submitted

\bibitem[\protect\astroncite{Wilms et~al.}{1998b}]{wilms:98b}
Wilms J., Nowak M.A., Pottschmidt K., et~al., 1998b, ApJ in preparation

\end{thebibliography}
\end{document}